\documentclass[aps,prl,reprint,floatfix]{revtex4-1}
\usepackage{graphicx,natbib}
\usepackage[bookmarks=false]{hyperref}
\usepackage{bm}
\begin{document}
\title{Spreading dynamics of infectious diseases on structured society with daily cycles}
\author{Kenichi Nakazato}
\affiliation{Bosch Center for Artificial intelligence, Corporate Research, Bosch Corporation, Tokyo, Japan, 150-8360}
\email{Kenichi.Nakazato@jp.bosch.com}
\author{Masanori Takano}
\affiliation{Akihabara Laboratory, CyberAgent, Inc., Tokyo, Japan, 101-0021}
\affiliation{Graduate School of Enginnering, The University of Tokyo, Tokyo, Japan, 113-8656}

\begin{abstract}
We are facing a common serious issue, infectious diseases, and trying to suppress the spreading of infection.
We need less contact with each other to decrease the chance of infection, but this means loss of economic activity, as well.
This tradeoff is inevitable in our society, because we still need direct communication and commuting, so far.
The focus of our paper is the structure of society, on which we have direct contacts.
We study on spreading process with artificial sosiety model, where each agent has daily cycle and go office and back home, every day.
At the same time, infection spreads along SIR model.
We show both slow infection and short commuting can be realized with some structures and vice versa.
The most effective factor for such features is modularity of society.
In highly modular society, agents live around the destined office, but agents commute long way to their office and can be infected fast, in not modular society.
The first infection point is one more factor for the features.
If the first infection takes place around the office, infection spreads slower.
On the contrary, if the first one takes place far away from the office, infection can be fast.
We show a design principle, high modularity and sparsely distributed offices, for good society and discuss on possible solutions in real society, where we live in.
\end{abstract}
\maketitle

\section{introduction}
Infectious diseases, like COVID-19, are one of our most concerning issues in recent days.
In our human society, infectious diseases are conveyed through human body, 
since many of them can not survive for long time out of human bodies. 
When humans are neighboring, they can easily infect from each host into others, through the air, droplet or direct contacts.
This means the spreading process of the infectious diseases can heavily depend not only on our immune system, but also on human social activity and structure.
Actually, many governments recommend less human migration or social gathering to decrease the potential chances of infection. 
On the other hand, the risk of massive infection is emphasized in the modern human society, because of change in our lifestyle, e.g. much higher mobility and wider radius of action.
As a result of such wider range of our activity, we have more chances to hit unknown infectious pathogens or other hosts.

In the recent situation with COVID-19, one of the central issue for us is the tradeoff between maintenance of human daily economy and decrease of infection chances.
We should work and iterate daily life cycles for our survival in economical meanings, while we want to decrease the chances to hit some other hosts and meetings, as a potential risk for infection, therefore.
To solve this seemingly incompatible desires at once, we study on the possibile structures of human society using an artificial society model here.
In the artificial society, we should commute to the office everyday. Here, as the commuting, we assume railways and its network.
In overpopulated cities, we mainly rely on railways and such an overpopulation itself can be a major risk for infection.

What types of configuration of cities take longer time for spreading or not?
If the spreading proceeds at once, hospitals can not afford satisfactory medical treatment and this can result in more victims.
In this meaning, we call a city with slow spreading as safe one and fast spreading as dangerous.
In daily lives of agents in our artificial society, they can meet other host agents in commutation, office and their home.
Since the many chances of meeting hosts means fast spreading, intuitively, 
agents will iterate daily life cycles without unnecessary hittings to other hosts, in safer cities.
Here we show results of variety of configurations and some design principles for safe or dangerous cities.

How can we model spreading dynamics itself?
The spreading process of infectious diseases have been studied intensely in the field of complex networks\citep{SIR_SF,epiSW,epi1,epiDegD,epiAda,NewmanMEJ}. 
Many of them adopt a kind of SIR model, which have sucseptible, infected and recovered state of agents, because of its easiness for understanding and implementation.
Along this line, we adopt SIR model as spreading dynamics in our artificial society, as well.

In the next section, we will introduce our artificial society and its formulation. 
Then spreading features are shown. 
Then, we show some design principles of safe and dangerous societies from analysis of the safe and dangerous social structures.
Finally, we discuss on impacts of social structures on infectious diseases and some possibile ideal structures for the future society.

\section{model}
In our artificial society, each agent lives in one of $N$ nodes. In each node, $U$ units of agents are living in it.
They have their own daily schedule unit by unit. Along the schedule, agents go to their office and come back home, every day.
In the society, there are $M$ office nodes.
To commute to each office, each agent should move along the paths between nodes.
Those paths between nodes mean railway and the network of nodes and paths is the structure of society.
As the basic configuration of our artificial society, we adopt circular network and there are $S$ short cut railway links.
Agents can move all of single paths with a unit time. Short cut links can mean express, therefore.
Needless to say, this is a version of known small world networks \citep{WS,Klein,NewmanSW,train-net}.

In each schedule, agents have flexibility on start and end time.
However, agents should be at work at noon, in the pre-determined office, and stay at the office predetermined duration, here we adopted one-third of a day.
In this meaning, they have flexibility on working hours.
To set the schedule, we can set start time along normal distribution, ${\cal N}(\mu,\sigma_M)$, where $\mu$ is the standard start time, $1/6$ of a day before noon and $\sigma_M$ is the standard deviation.
Then end time can be set to be satisfied with the worktime, $1/3$ of a day.

In the schedule, each agent should have only one office.
The office can be selected along the path length, $L_{ij}$, from home node, $i$, to the office node, $j$.
From all possible paths, from the home node to every office nodes, each agent can select only one path along the weight, $\exp(-\beta L_{ij})$.
Each agent iterates own daily cycle along the schedule, including commutation path and office, everyday.

To study the spreading process, we firstly choice a node, other than offices, randomly and put a little number of infected agents there.
The other agents hit such infected ones and can be infected, if they are at the same node or railway.
To be noted, each railway has direction and infection can occur between units on the same direction.
We assumue the infection can occur in mean field way, that is infection probability is proportional to the ratio of infected agents there.
At the same time, infected ones can be recovered at a constant ratio.
This dynamics can be expressed in a variation of known SIR equations,
\begin{eqnarray}
\frac{dS_{ij}}{dt}&=&-a \bar{I}S_{ij},\\
\frac{dI_{ij}}{dt}&=&a I_{ij}\bar{S}-b I_{ij},\\
\frac{dR_{ij}}{dt}&=&b I_{ij}.
\end{eqnarray}

In these equations, each state ratio, $S_{ij}$, $I_{ij}$ and $R_{ij}$, are the one with home node $i$ and unit $j$.
Mean values, $\bar{I}$ and $\bar{S}$, are averaged ones on all units at the same site or railway at each time.
Parameters, $a$ and $b$, are constant.

Along this dynamics, all units of agents change their own ratio between, $S$, $I$ and $R$, in each time step.
In reality, there should be a physical limit of capacity for railways, but we ignore such a limit for simplicity.
In addition, we set the same population for all of nodes for simplicity, as well.

We can focus on just the effect of social structure on the spreading process with our artificial society.

\section{results}
\subsection{random generated configurations}
We studied on randomly generated configurations, firstly.
Here we show some results with the number of nodes, $N=10$, the number of offices, $M=2$, and the number of short cuts, $S=3$.
In tests, we randomly determined the place of offices and short cuts test by test.
A day has $90$ unit times and worktime is $30$, therefore. In this setting and standard schedule, agents work from 30 to 60 in unit time.
But, actually, all schedules are determined with standard deviation, $\sigma_M=10$, and path selection parameter,$\beta=1.0$.
SIR parameters are set, $a=0.1$ and $b=0.1$.

\begin{figure}[htb]
\center{\includegraphics[bb=0 0 320 240, width=0.5\textwidth]{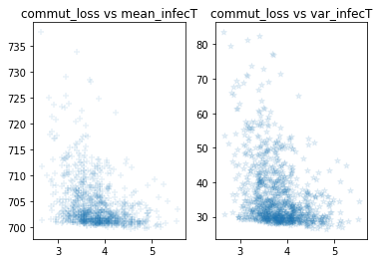}}
\caption{Infection and commuting time. The left plot is commuting loss vs mean infection time.
The right one is commuting loss vs standard deviation of infection time. 
Commuting loss is the time for commuting averaged over all agents in the society.
Infection time is evaluated with that of all nodes in the society.}
\label{fig:XXX}
\end{figure}

We show scatter plots in FIG.\ref{fig:XXX}, where the horizontal axis is commuting time and vertical axis is infection time.
As commuting time, we calculated mean value of one way commuting unit times.
Here we call it as commuting loss.
The infection time is defined as mean unit time till the all of the infected ratio, $\sum_jI_{ij}/U$, living in node $j$, become larger than a threshold, $0.3$.
If recovery rate, $R$, is too high, max of infection ratio can turn into decline before the threshould.
We set the threshould so that each infection ratio can achieve in the course of tests.
In the second plot, we show the scatter plot with the standard deviation of infection time as vertical axis.

We can confirm two significant features in them.
Firstly, some societies can achieve both little commuting loss and slow infection, low infection time, at the same time.
On the contrary, some societies have both high commuting loss and fast infection at the same time.
From these two aspects, safe and convenient society can be achievable and vice versa.

\begin{figure}[htb]
\center{\includegraphics[bb=0 0 720 360,width=0.5\textwidth]{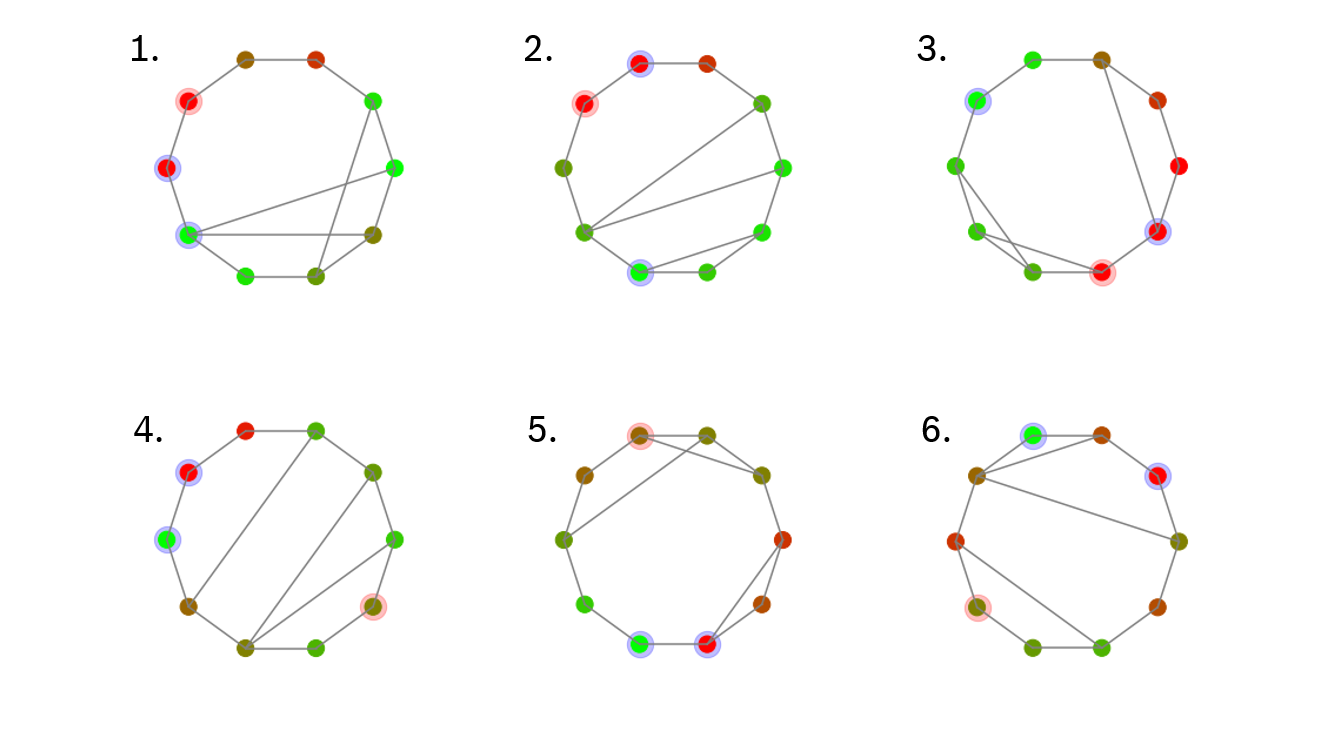}}
\caption{Top and worst societies. The top three, slow infection, societies are $1$, $2$ and $3$.
The worst, fast infection, ones are $4$, $5$ and $6$.
Those are selected from randomly generated $1000$ societies along the infection time.}
\label{fig:YYY}
\end{figure}

Then what types of societies have such favorable and unfavorable features?
Here we show some of the top favorable and the worst favorable ones, in FIG.\ref{fig:YYY}.
We selected them from $1000$ tests with the same parameter settings, along the infection time.
We colored nodes along the ratio of destinations of agents living there.
Office nodes have outer blue disk as well and the first infected node, the source of infection, is colored with outer red disk.
If many of agents of a node go to the office with red color, the node have similar color.
If agents, living in a same node, go variety of office nodes, the node have mixed color.
At a first glance, we can notice higher contrast in top societies and lower contrast in worst ones.
Once a node is infected, infection ratio can grow within the same colored cluster, because many agents living there share long time at the office node.
Then infection spreads into the other cluster through some agents with other offices.
Since mixing of infected and susceptible agents can easily occur in the nodes with variety of destination offices, mixed colored nodes can have shorter infected time.
On the other hand, intuitively, we can guess high correlation between fast infection and average shortest path length.

\begin{figure}[htb]
\center{\includegraphics[bb=0 0 480 280,width=0.5\textwidth]{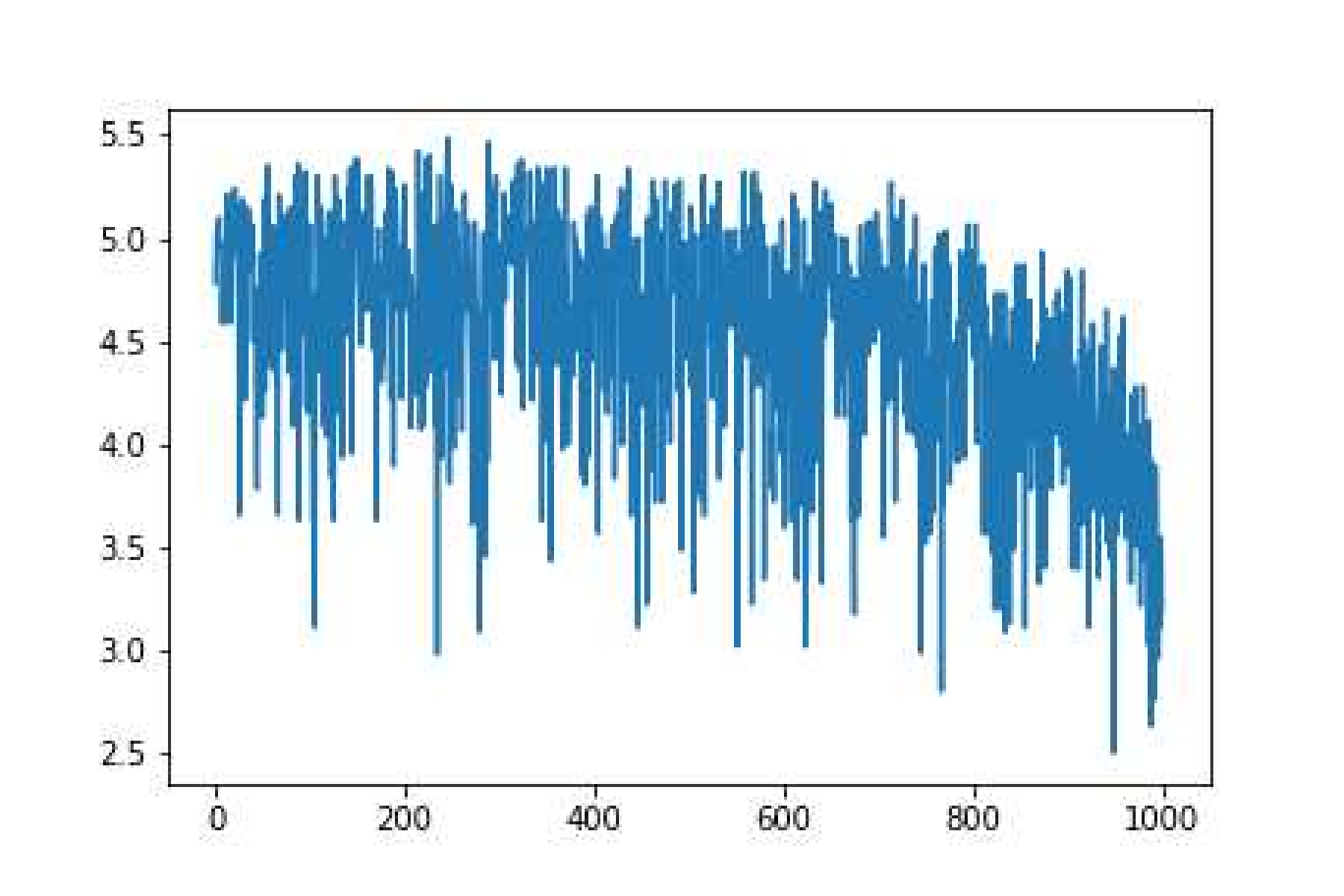}}
\caption{Entropy of destinations. Each agent has different office, destination, to go every day.
Here we calculated entropy of the distribution of the destinations and plotted along the ranking of infection time.}
\label{fig:ZZZ}
\end{figure}

\begin{figure}[htb]
\center{\includegraphics[bb=0 0 480 320,width=0.5\textwidth]{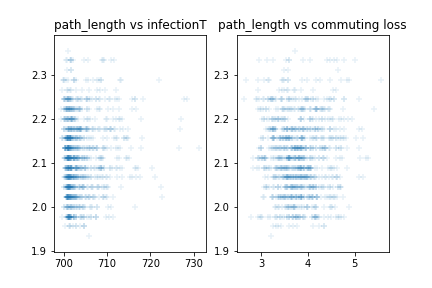}}
\caption{Correlation between path length and infection or commuting.
Horizontal axes are mean infection time, in the left, and commuting time, in the right.
Vertical asix is average shortest paths length of each society.
}
\label{fig:AAA}
\end{figure}

We can confirm these hypotheses in the next, FIG.\ref{fig:ZZZ} and FIG.\ref{fig:AAA}.
In FIG.\ref{fig:ZZZ}, we show entropy of destinations. Since agents have different pre-determined schedules and offices unit by unit, we can defne destination entropy for each node.
For each node, $i$, the destination ratio, $D_{ij}$, can be defined along the destination office, $j$.
If almost all of agents in a node go to the same office, the destination ratio, $D_{ij}$, must be biased.
This means entropy of the destination ratio, $-\sum_jD_{ij}\log{D_{ij}}$, shows low value.
In FIG.\ref{fig:ZZZ}, the sum of destination entropy, $-\sum_{ij}D_{ij}\log{D_{ij}}$, is plotted along the rank of infection time.
As we can confirm, destination entropy decreases along the rank, in other words, small destination entropy can be a sign of slow infection.

In general, information propagation is fast on the network with small average shortest path length \citep{masuda}.
As we mentioned above, this leads us one more hypothesis on the correlation between average shortest path length and infection time.
In FIG.\ref{fig:AAA}, we show the correlation and one more, between average shortest path length and average commuting time, in parallel.
We can easily confirm no correlations there, contrary to our expectations.
In our society, infection spreads through mixing, between susceptible and infected agents, and such mixing occurs along the daily cycles.
If agents move in random manner, the process would be similar to diffusion on the social network, however agents in our society have regular daily schedules.
That would be the reason for no correlation between average shortest path length and infection or commuting time, different from diffusion process.

\begin{figure}[htb]
\center{\includegraphics[bb=0 0 480 280,width=0.5\textwidth]{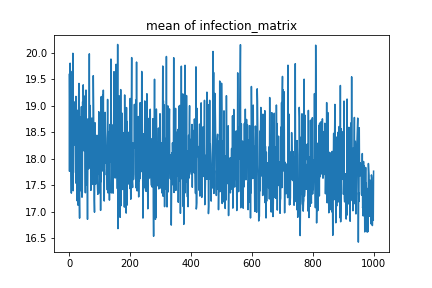}}
\caption{Effect of infection matrix. We defined infection matrix with the count of overlappings of daily commuting between different nodes.
As the effect of infection matrix, we plotted the mean value of the matrix along the infection time ranking.}
\label{fig:BBB}
\end{figure}

Our results, FIG.\ref{fig:ZZZ} and FIG.\ref{fig:AAA}, suggest clear separation of destinations among different nodes is the major key factor for safe and convenient society.
However, we can confirm strong deviations in the trend.
In the spreading process, infection can proceed within the same node easily, but not between different nodes.
We can evaluate the chance of infection between different nodes with overlapping of schedules.
Here we compare all combinations of scheduled paths and count such overlappings between different nodes, $i$ and $j$.
We show one more result on the count of overlappings, $C_{ij}$.
In FIG.\ref{fig:BBB}, mean of the count, $\sum_{ij}C_{ij}/N^2$, is plotted along the rank of infection time.
We can confirm weak declining tendency along the rank.
This means degree of overlappings in schedules can be a weak sign for fast infection.
However, we have strong deviation again.
In reality, further infection requires not only chances to infect nearest neighbors in the count matrix, $C_{ij}$, but also second or higher order neighbors in the matrix.
In such a later stage in spreading process, more detailed structure should be considered into estimation of infection time.

\subsection{effect of the first infection point}
\begin{figure}[htb]
\center{\includegraphics[bb=0 0 1800 1200,width=0.5\textwidth]{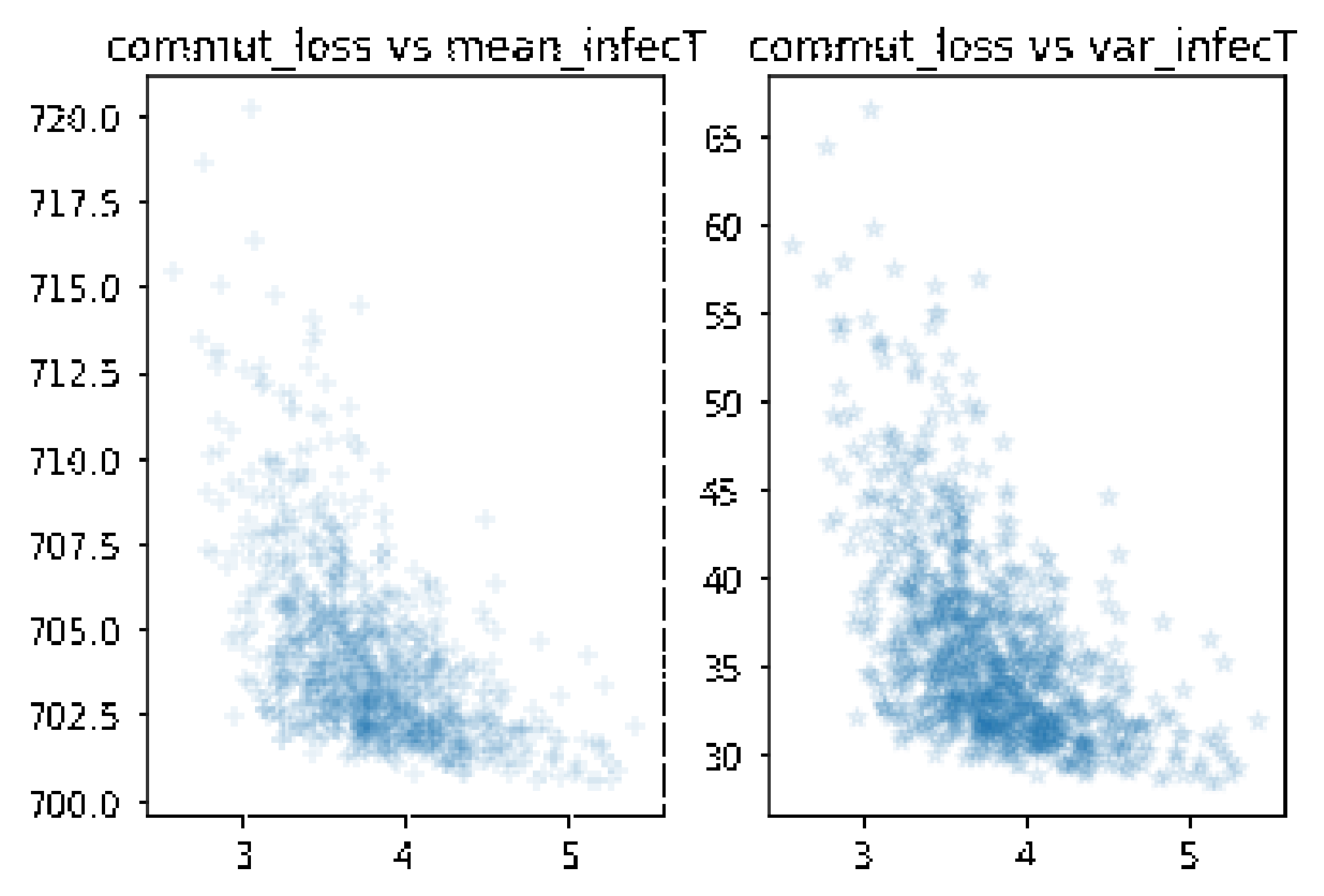}}
\caption{Infection time vs commuting loss. Scatter plot between mean infection time and commuting loss, in the left, and between standard deviation of infection time and commuting loss, in the right.
These plots are calculated by averaging over all possible initial infection points in each society.}
\label{fig:FFF}
\end{figure}

Here we focus on the reasons for the deviations in our results, FIG.\ref{fig:ZZZ} and FIG.\ref{fig:BBB}.
As the possibility, we show results on the effect of the place of first infection.
We tested all possible first infection places, other than office nodes for every societies.
Firstly, we show scatter plots on infection time and commuting loss, in the FIG.\ref{fig:FFF}
Here we used the same parameters in the previous test.
As we can see, we can easily confirm the same tendency in the previous test, except for the vertical scale.
In this result, deviation of infection time is smaller than the previous one.
Since we show the mean infection time averaged over all possible start infection nodes for every configurations, deviation can be decreased in the result.

\begin{figure}[htb]
\center{\includegraphics[bb=0 0 480 280,width=0.5\textwidth]{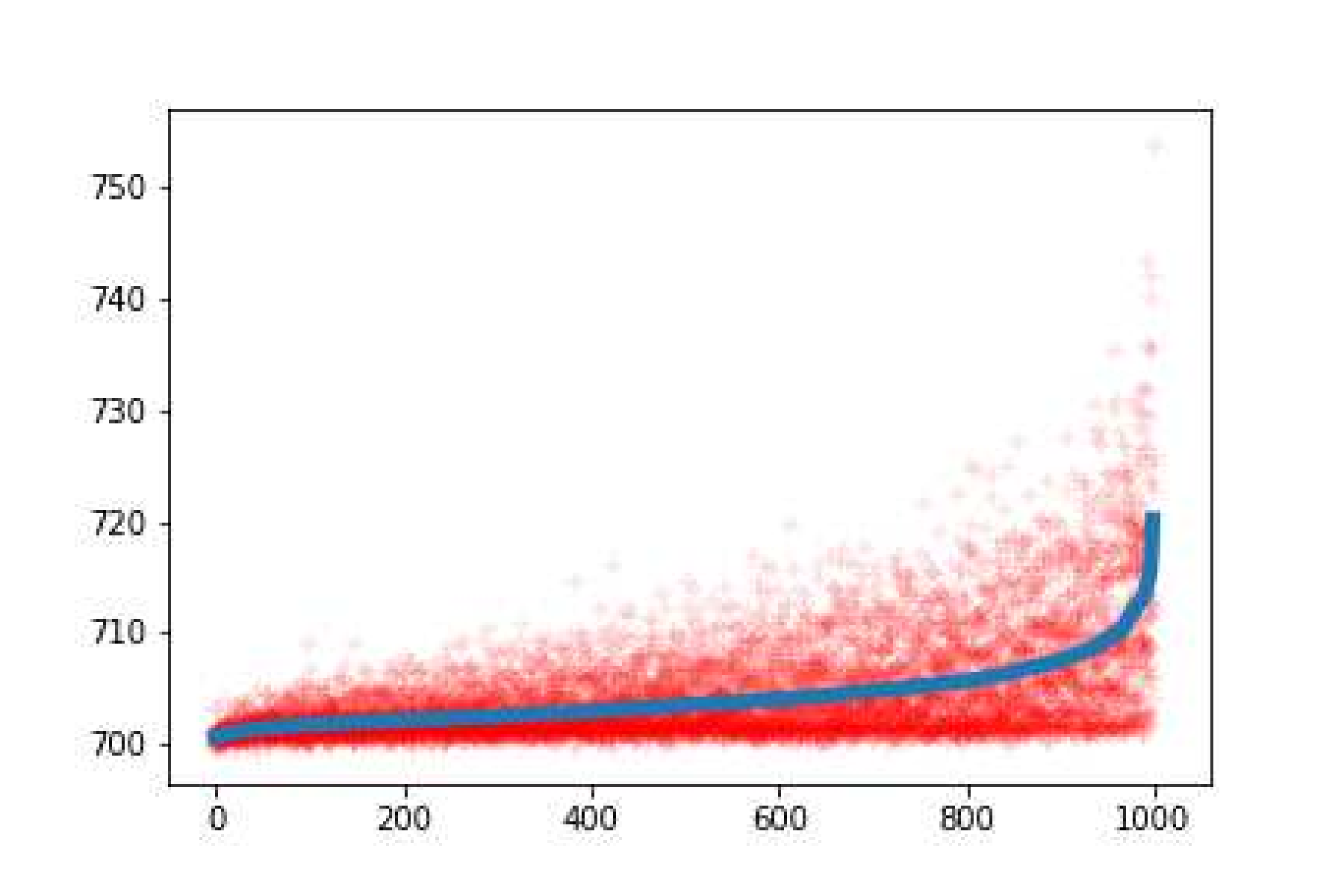}}
\caption{Infection time variation. Infection times are plotted, in red dots, for all possible cases of initial infection.
Horizontal axis is the rank of mean infection time. Vertical axis is infection time.
Mean infection time is also plotted in the blue curve.}
\label{fig:VVV}
\end{figure}

To confirm this, we show one more plot in the FIG.\ref{fig:VVV}.
In this plot, we show all of infection time along the mean infection time ranking.
The mean infection time, in FIG.\ref{fig:FFF}, is plotted in blue line, here.
In addition, we show all of infection time for each case in red dots.
Since each infection time is sampled from such a distribution, we see more deviated one in the FIG.\ref{fig:XXX}, consequently.
At the same time, this result tells us infection time can heavily depend on the place of first infection.

\begin{figure}[htb]
\center{\includegraphics[bb=0 0 480 280,width=0.5\textwidth]{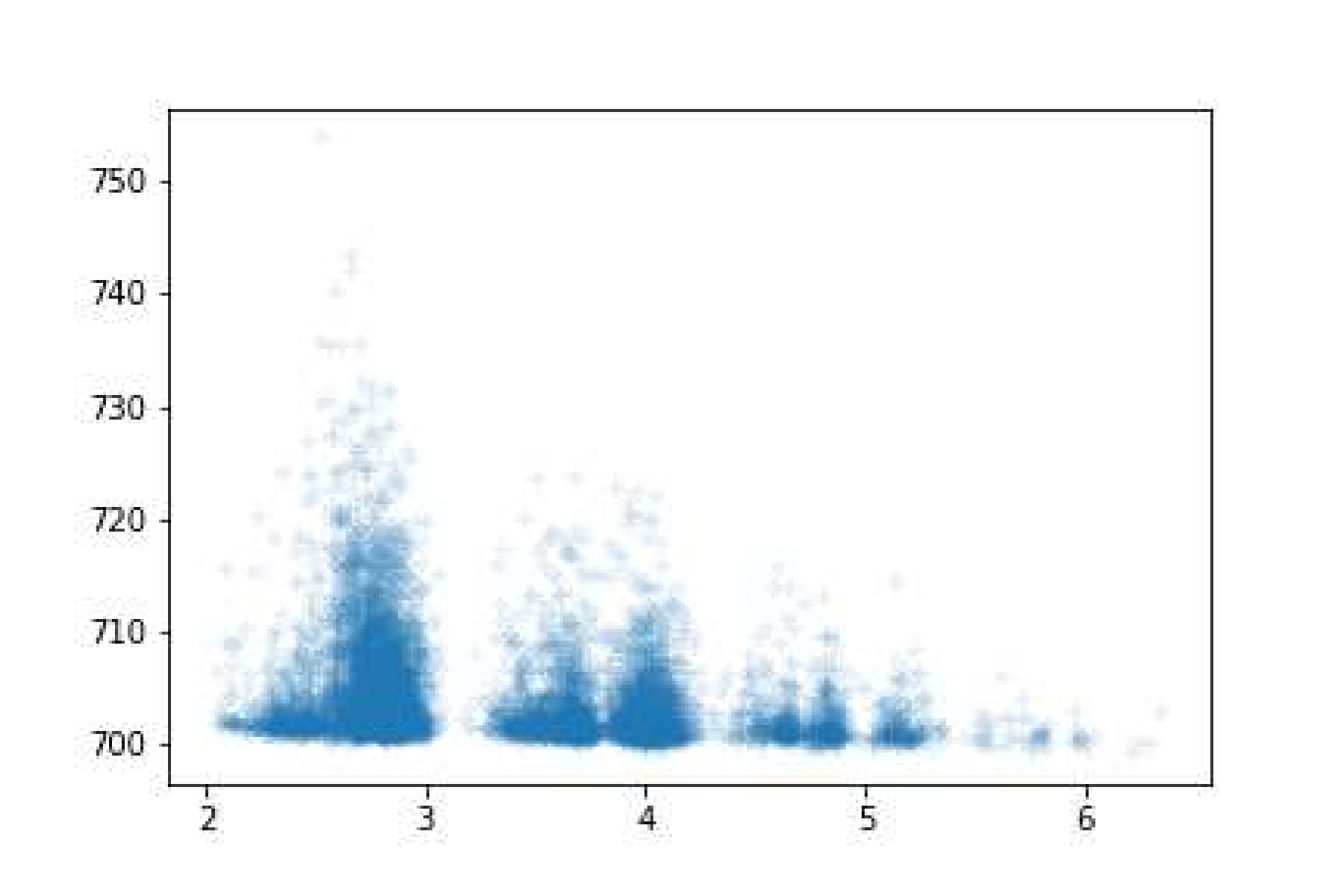}}
\caption{Correlation between commuting time from the first infection node and infection time.
Commuting time is the weighted average of the path lengths from the first infection node to every offices.
The weight is the same as selection weight in scheduling.}
\label{fig:III}
\end{figure}

How the place of first infection can bring about such difference in mean infection time?
We can guess the distance, between office and the first infection, may be the main factor of that.
We show the result of correlation between them in the FIG.\ref{fig:III}.
In the cases of shorter distance between office and the first infection, we can large deviation of infection time.
On the contrary, we can not see such large deviation in the cases of longer distance.
This result may be conter-intuitive, but we can consider the destination entropy to understand this.
Slow infection can be observed in low destination entropy cases.
In such a low destination entropy case, we can easily confirm clusters defined with the destinations.
If the first infection takes place near an office, spreading would consist of multi steps with some clusters and take long time.
On the contrary, it would not take so long time to spread, if the first infection takes place distant from offices.
In this case, the first infection can bring about simultaneous infection in different clusters.

\begin{figure}[htb]
\center{\includegraphics[bb=0 0 720 360,width=0.5\textwidth]{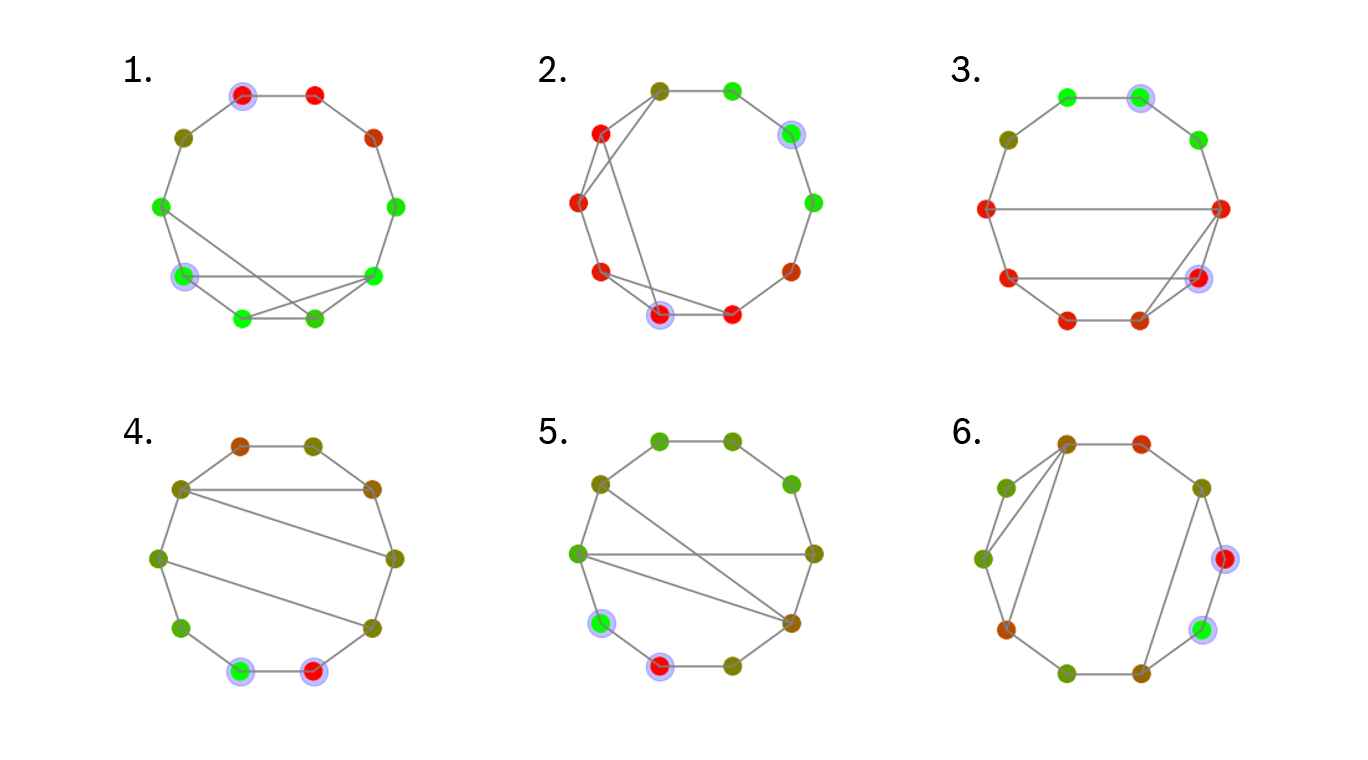}}
\caption{Top and worst societies against all possible initial infections. The top three, slow infection, societies are $1$, $2$ and $3$.
The worst, fast infection, ones are $4$, $5$ and $6$.
Those are selected from randomly generated $1000$ societies along the infection time.}
\label{fig:NNN}
\end{figure}

Then how about the structure of societies?
We show top and worst societies in FIG.\ref{fig:NNN}.
Here we selected top 3 societies and worst 3 societies along the mean infection time ranking.
As we can confirm, separation of destinations, office nodes, is clearer than the previous case, FIG.\ref{fig:YYY}.

\subsection{variation of societies}
We here test more variety of possibilities.
The results shown here are calculated with other parameter sets with more nodes, offices and shortcuts.
Other parameters are the same one as the previous tests.

\begin{figure}[htb]
\center{\includegraphics[bb=0 0 480 320,width=0.5\textwidth]{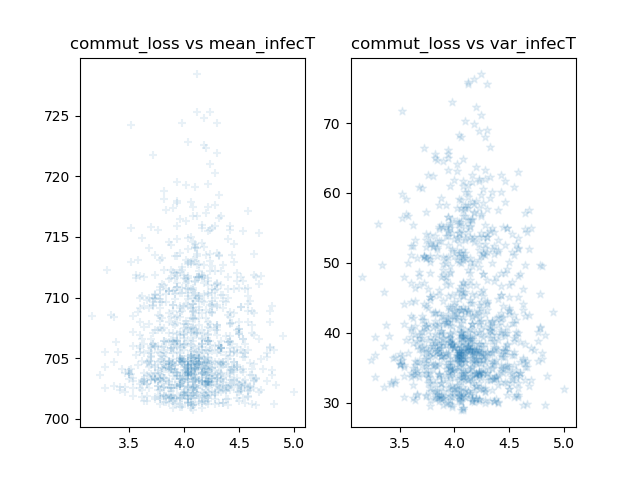}}
\caption{Infection time vs commuting loss with a case of many shortcuts.
Scatter plot between mean infection time and commuting loss, in the left, and between standard deviation of infection time and commuting loss, in the right.
This is the result with parameter set, $N=10$, $M=4$ and $S=10$.}
\label{fig:VSUM1}
\end{figure}

Totally, we can confirm the same tendency in the correlation between infection time and commuting loss, 
however we found some defferent features at the same time.
If we increase the shortcuts, then the tendency was lost gradually, in FIG.\ref{fig:VSUM1}.
Alternatevily, we see high deviation of infection time at the mid commuting loss.

\begin{figure}[htb]
\center{\includegraphics[bb=0 0 480 320,width=0.5\textwidth]{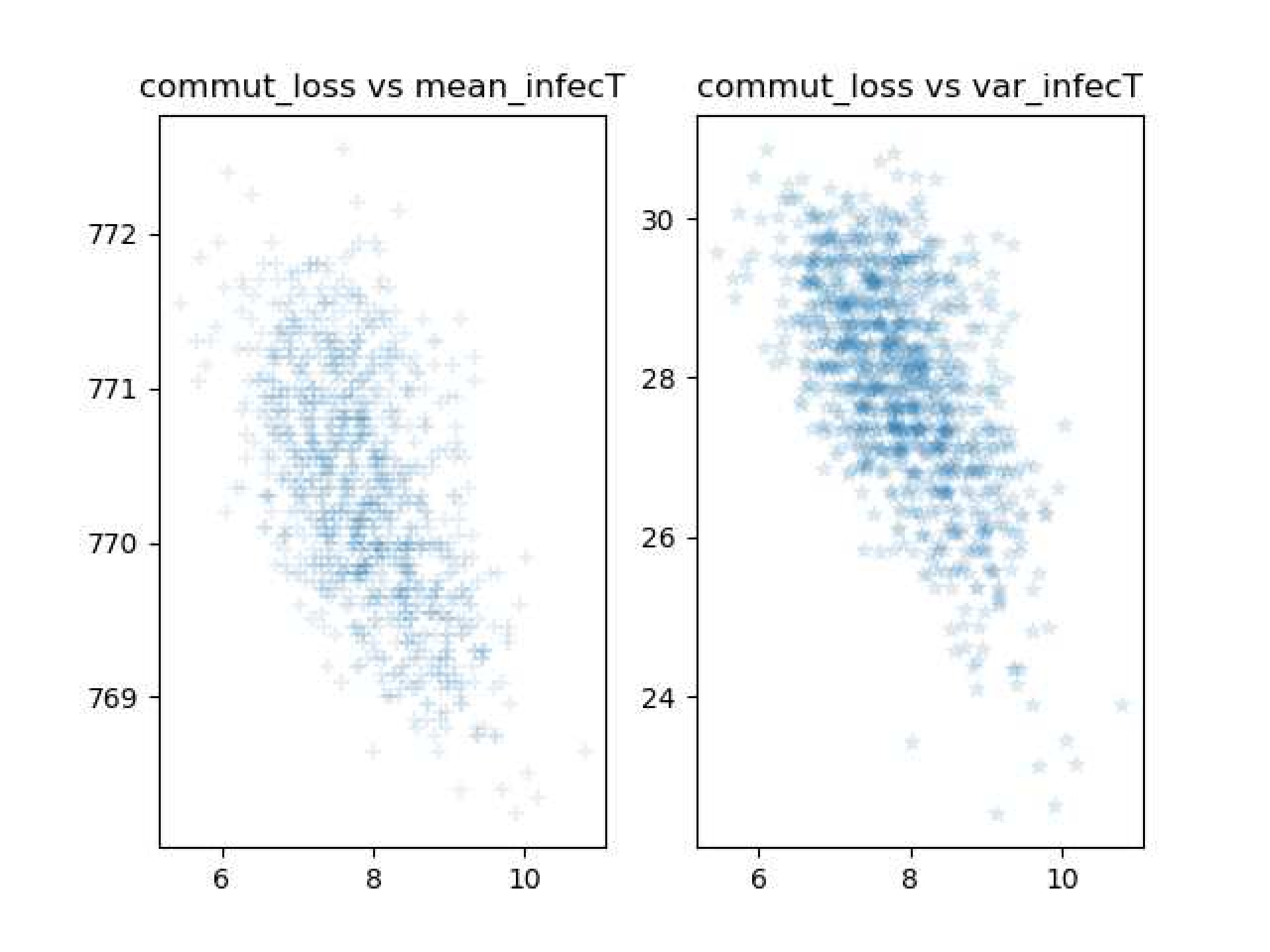}}
\caption{Infection time vs commuting loss with only one office.
Scatter plot between mean infection time and commuting loss, in the left, and between standard deviation of infection time and commuting loss, in the right.
This is the result with parameter set, $N=20$, $M=1$ and $S=12$.}
\label{fig:VSUM2}
\end{figure}

One more different feature is shown in FIG.\ref{fig:VSUM2}.
A society with only one office node and some shortcuts has many configurations for safety rather than others.
We can easily confirm more samples in the region of small commuting loss and slow infection.
However, this seems to be a singular with the case of only one office.
We could not find this type of scatter plot in other cases with more offices.

\begin{figure}[htb]
\center{\includegraphics[bb=0 0 720 360,width=0.5\textwidth]{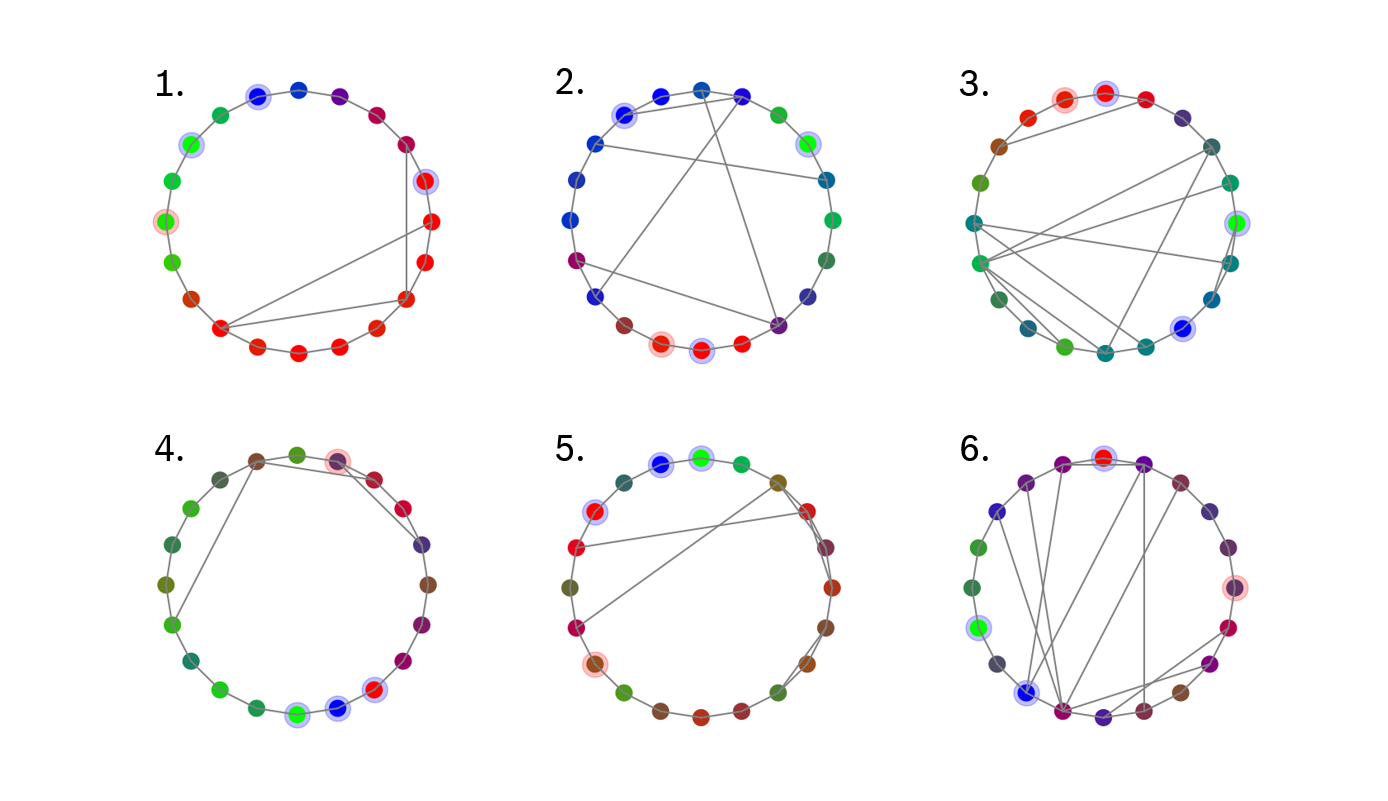}}
\caption{Top and worst societies with 20 nodes and 3 offices. The top three, slow infection, societies are $1$, $2$ and $3$.
The worst, fast infection, ones are $4$, $5$ and $6$.
Those are selected from randomly generated $1000$ societies along the infection time.}
\label{fig:VSOC1}
\end{figure}

Next, we show the features same as the previous tests.
In FIG.\ref{fig:VSOC1}, we show structures of society with variety of parameters.
As the same as the previous ones, FIG.\ref{fig:BBB} and FIG.\ref{fig:NNN} , we can confirm clear separation of destinations as top ranking societies.
On the contrary, we can confirm mixed destinations in the worst ranking societies, again.

\begin{figure}[htb]
\center{\includegraphics[bb=0 0 720 360,width=0.5\textwidth]{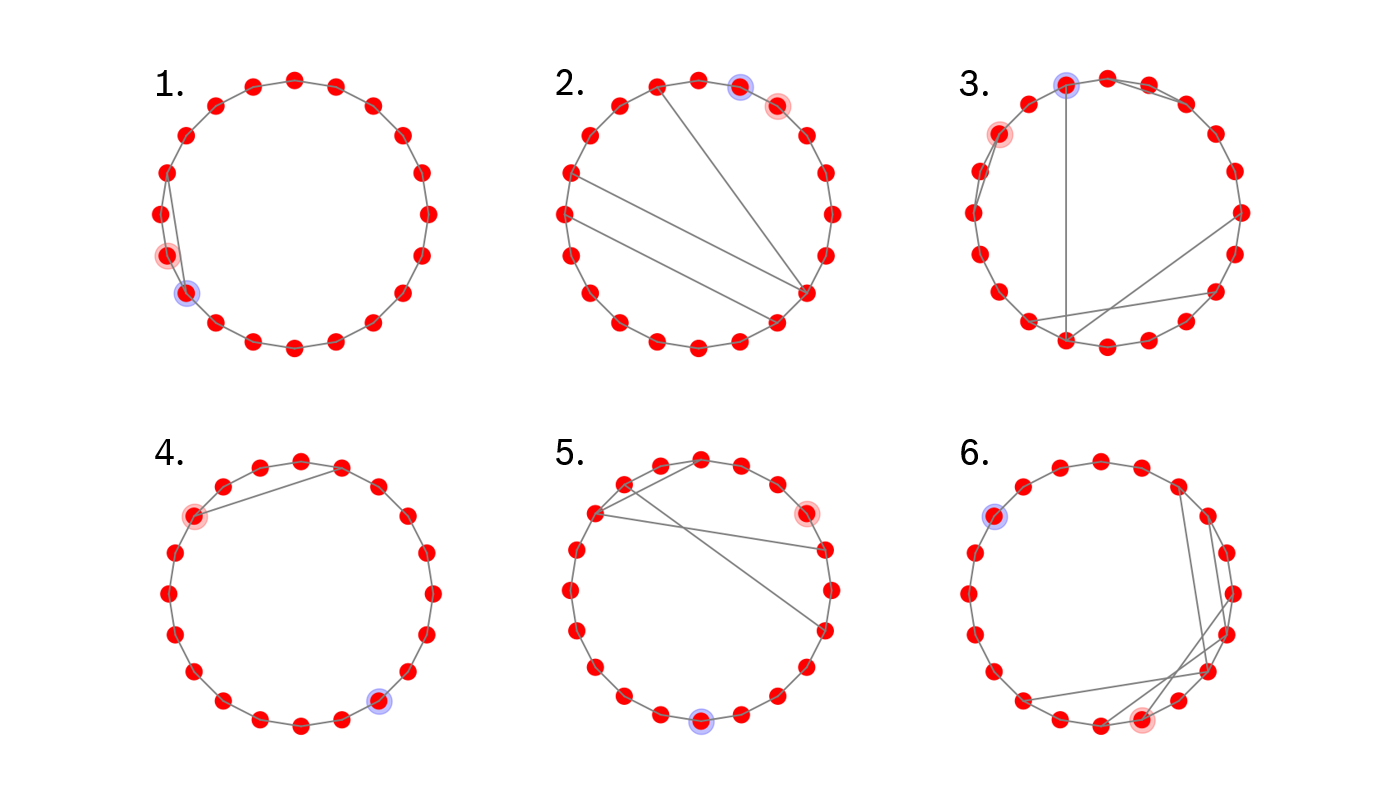}}
\caption{Top and worst societies with only one office. The top three, slow infection, societies are $1$, $2$ and $3$.
The worst, fast infection, ones are $4$, $5$ and $6$.
Those are selected from randomly generated $1000$ societies along the infection time.}
\label{fig:VSOC2}
\end{figure}

We show one more specific cases with only one office.
In FIG.\ref{fig:VSOC2}, we show structures of societies.
Here, we can confirm safe societies have shortcuts on the places, where more agents can use for commuting.
On the contrary, shortcuts are not convenient in the case of high commuting loss and fast infection.
In the case of only one office node, almost all agents share similar working time at the same office.
In this case, small commuting loss means slow infection in straightforward.

\subsection{design principles for good/bad societies}
\begin{figure}[htb]
\center{\includegraphics[bb=0 0 480 280,width=0.5\textwidth]{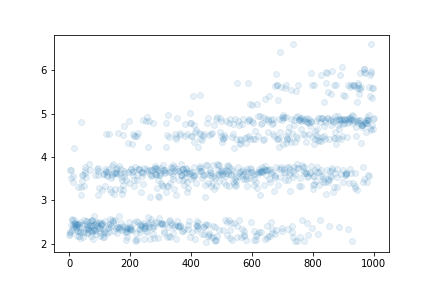}}
\caption{Office sparsity.
As an index for office sparsity, the distance between offices in each society is plotted along the rank of infection time.
As the distance, we calculated weighted average over all possible simple paths between offices in each society.
The weight is determined in the same way as selection of paths in scheduling.
Here we used results of all possible first infection nodes test.}
\label{fig:OOO}
\end{figure}

Let's go back to the original interest of us.
Here we investigate on design principles for good/bad societies.
In our definition, good society means slow infection and short commuting.
Bad society means fast infection and long commuting.
We studied on some factors, which can result in good or bad societies in the previous sections.
One is the destination entropy.
We confirmed the destination entropy may be one of the main factors, in FIG.\ref{fig:BBB}, FIG.\ref{fig:NNN} and FIG.\ref{fig:ZZZ}.
At the same time, we notice one more feature in FIG.\ref{fig:BBB} and FIG.\ref{fig:NNN}.
In those societies, we see densely placed office nodes in bad ones and sparsely placed ones in good ones.
To confirm this tendency, we show mean distance between office nodes along the infection time ranking.
In FIG.\ref{fig:OOO}, we can confirm short distance means fast infection and long distance means slow infection.
This result leads us the possibility of one more main factor for good/bad societies.

\begin{figure}[htb]
\center{\includegraphics[bb=0 0 1800 1200,width=0.5\textwidth]{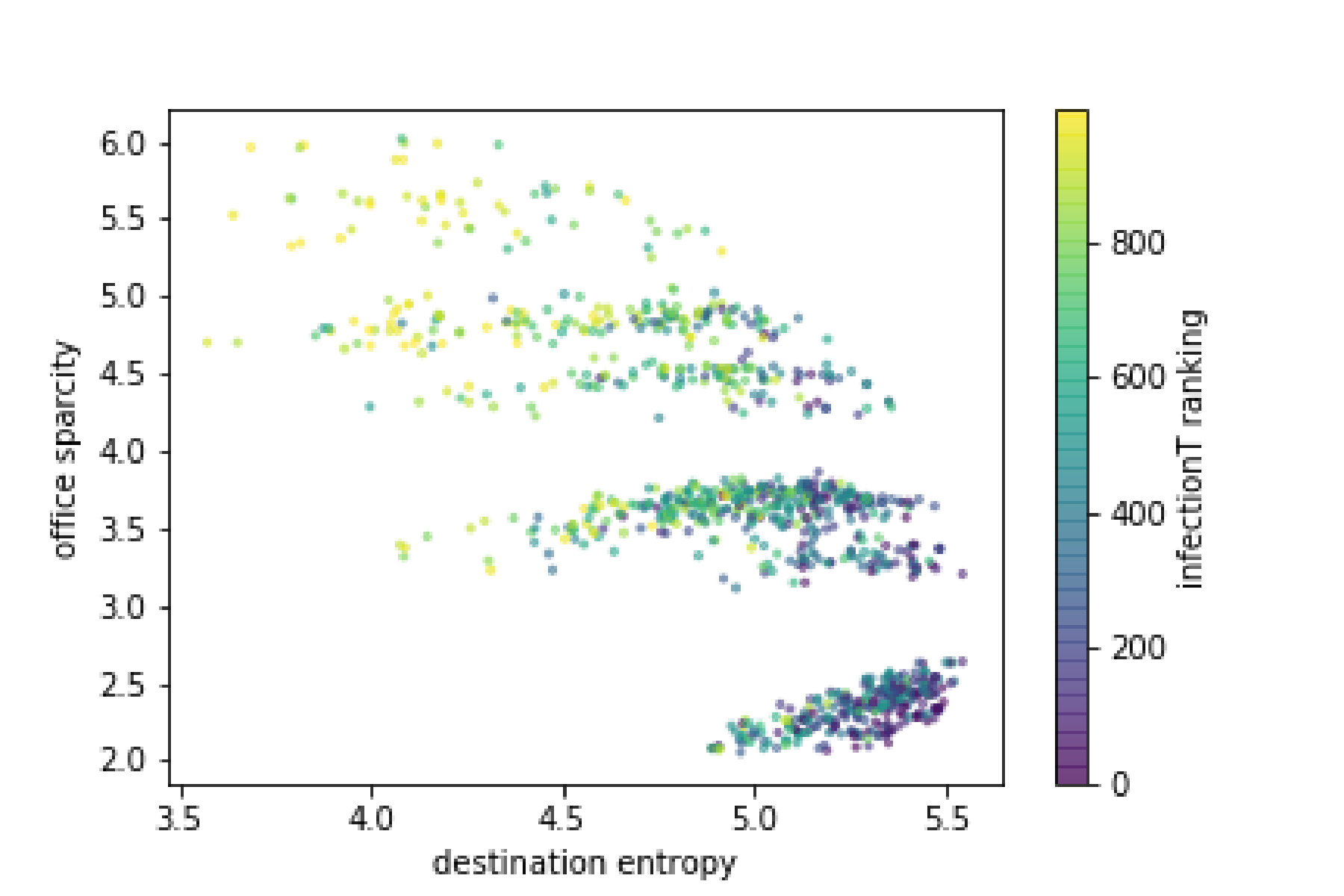}}
\caption{Performance diagram. Horizontal axis is destination entropy.
Vertical axis is office sparcity. Color is infection time ranking.
High infection time ranking means slow infection.
Here we used results of all possible first infection nodes test.}
\label{fig:PPP}
\end{figure}

Then which one is the most effective for good/bad societies?
We can confirm this with a performance diagram, in FIG.\ref{fig:PPP}.
We show correlation between destination entropy, office sparcity and infection time ranking.
As we can easily confirm, high destination entropy means fast infection time.
At the same time, low office sparcity means fast infection, as well.
There are weak negative correlation between the two factors, but we expect slow infection if both of them are satisfied.

\section{discussion}
In this paper, we are studying on the possible configurations of social structure.
Especially, we focus on daily cycles, including commuting, of agents in the society.
In the facing of serious infectious diseases, like COVID-19, it is not easy to prevent infection itself, partially because unknown features of the pathgen.
However, the pathogen is conveyed mainly by human and we can manage the spreading process in some ways, therefore.
Ideally we should prevent contact with other persons to decrease the chance of infection in principle.
However we should keep our daily cycles to maintain our economic condition, at the same time, and can not escape from contacts with other persons.
In this meaning, we have tradeoff between economic and survival activity.
That is why we focus on the social structure, where humans iterate own daily economic activity and contact with each other.
Here we are assuming railways as commuting method and the society is a network connected with the railways.

In this setting, our ideal society should have the features, slow infection and short commuting time.
Can we have possibility for such an ideal society? The answer is yes, in our context, FIG.\ref{fig:XXX} and FIG.\ref{fig:FFF}.
However, we have the other possibility for nonideal extreme, fast infection and long commuting time, at the same time.
In ideal societies, agents show clear segragation in the meaning of destined offices in their daily cicles, FIG.\ref{fig:ZZZ}
One more stliking feature of such a society is sparsely distributed offices, FIG.\ref{fig:OOO}.
Since agents share long time mainly within the same destined ones, such clear segragation means fast infection in the same group.
However, if we see the other groups, with other destined offices, they have only limited chances for infection from the first infected group.
At the same time, this type of social structure means short commuting time because of the modular structure \citep{mod-net,mod-diff}.
Actually, we can say both of features are satisfied with in many cases, FIG.\ref{fig:PPP}.

To be noted, we can not miss the significant effect of the place of first infection on infection time, FIG.\ref{fig:VVV} and FIG.\ref{fig:III}.
Surprisingly, our results suggest more risk in the case of the first infection at distant place from offices.
On the contrary, if the first infection occur around offices, infection can take longer time.
We can explain the reason of the difference between the two cases with the possibility of parallel spreadings.
In the case of first infection around offices, infection spreads from a group to other group step by step.
However, in the other case, infection can spread in multiple groups at the same time.

We assumed the same infection parameters in the home, office and railways, but it should heavily depend on population density.
If we can assume the setting, without the effect of population density, we can depict the spreading process as the iterative mixing of sucseptible and infected ones at the office and home, in short.
In this situation, the solution for ideal society is modular separated structure in the meaning of destined offices.
However, if we can assume overpopulation at railways, like Tokyo, we can not ignore the significance of infection on the railways.
If agents with variety of destined offices share the same railways, the mixing, at the railways, will have more significant impact on spreading process.
Again, we can expect modularity would be the solution in such a case.

Needless to say, our model can help to study on strategies for infectious desease\citep{Estrategy}, like the effect of shutting down of society for suppressing infection.
We do not have space any more to report on it, but we can add some options for rescheduling the daily cycles.
We can test some possible shutting down strategies for variety of purposes, like fast recovery or suppression of the maximum level of infection with minimum duration of shutting downs.
Since shutting down always means economic loss, we hope we can find best strategies fit to each society with given structure.

Even if the solution is the modularity, we can not escape from the fast infection within the first infected group, sharing the same office, in the society.
The ultimate solution would be remote work at home or neighboring spaces without long commuting, therefore.
This solution can be applied for societies with densely placed offices, without modularity.
Since we are living in the era with enhanced infrastructure for remote communication, it is not hard task, we believe.
In addition, we have one more possibility with internet of things, which will decrease the necessity of physical operations on site.
With the aid of connected world, we can decrease the necessity of daily commuting itself.
On the other hand, even if we have enhanced infrastructure for remote communication, we still do not have enough technologies for natural communication, e.g. virtual reality for non-verval information facilitating \citep{takano}.
Technologies, for both virtual and physical space, should be synergistically enhanced, with the study of possible spaces, for our healthy and happy life.

\footnotesize
\bibliographystyle{apalike}
\bibliography{isoc_paps}

\begin{thebibliography}{}

\bibitem[Burke and Chakravarty, 2006]{Estrategy}
Burke, DS.~Epstein, J. C. D. P. J. C. K. S.~R. and Chakravarty, S. (2006).
\newblock Individual-based computational modeling of smallpox epidemic control
  strategies.
\newblock {\em Acad Emerg med.}, 13:1142--9.

\bibitem[Gross and Blasius, 2006]{epiAda}
Gross, T.~D'Lima, C. and Blasius, B. (2006).
\newblock Epidemic dynamics on an adaptive network.
\newblock {\em Physical Review Letters}, 96:208701.

\bibitem[Han, 2007]{epiSW}
Han, H. (2007).
\newblock Desease spreading with epidemic alert on small-world networks.
\newblock {\em Physics Letters A}, 365:1--5.

\bibitem[Kleinberg, 2000]{Klein}
Kleinberg, J. (2000).
\newblock Navigation in a small world.
\newblock {\em nature}, 406:845.

\bibitem[Masuda and Lambiotte, 2017]{masuda}
Masuda, N.~Porter, M. and Lambiotte, R. (2017).
\newblock Random walks and diffusion on networks.
\newblock {\em Physics Reports}, 716-717:1.

\bibitem[Moore and Newman, 2000]{NewmanMEJ}
Moore, C. and Newman, M. (2000).
\newblock Epidemics and percolation in small-world networks.
\newblock {\em Physical Review E}, 61:5678.

\bibitem[Nadini and Perra, 2018]{mod-net}
Nadini, M.~Sun, K. U. E. S. M. R.~A. and Perra, N. (2018).
\newblock Epidemic spreading in modular time-varying networks.
\newblock {\em Scientific reports}, 8:2352.

\bibitem[Nematzadeh and Ahn, 2014]{mod-diff}
Nematzadeh, A.~Ferrara, E. F.~A. and Ahn, Y. (2014).
\newblock Optimal network modularity for information diffusion.
\newblock {\em Physical Review Letters}, 113:088701.

\bibitem[Newman MEJ.~Moore and Watts, 2000]{NewmanSW}
Newman MEJ.~Moore, C. and Watts, D. (2000).
\newblock Mean-field solution of the small-world network model.
\newblock {\em Physical Review Letters}, 84:3201.

\bibitem[Pastor-Satorras and Vespignani, 2001]{epi1}
Pastor-Satorras, R. and Vespignani, A. (2001).
\newblock Epidemic dynamics and endemic states in complex networks.
\newblock {\em Physical Review E}, 63:066117.

\bibitem[Seaton and Hackett, 2004]{train-net}
Seaton, K. and Hackett, L. (2004).
\newblock Stations, trains and small-world networks.
\newblock {\em Physica A}, 339:635--644.

\bibitem[Takano and Tsunoda, 2019]{takano}
Takano, M. and Tsunoda, T. (2019).
\newblock Self-disclosure of bullied-experiences and social support in avatar
  communication.
\newblock {\em Proc. of 13th international conference on web and social media},
  13.

\bibitem[Wang and Liu, 2014]{epiDegD}
Wang, W.~Tang, M. Z. H. G. H. D.~Y. and Liu, Z. (2014).
\newblock Epidemic spreading on complex networks with general degree and weight
  distributions.
\newblock {\em Physical Review E}, 90:042803.

\bibitem[Watts and Strogatz, 1998]{WS}
Watts, D. and Strogatz, S. (1998).
\newblock Collective dynamics of small-world networks.
\newblock {\em nature}, 393:440--442.

\bibitem[Zhou and Wang, 2006]{SIR_SF}
Zhou, T.~Liu, J. B. W. C.~G. and Wang, B. (2006).
\newblock Behaviors of susceptible-infected epidemics on scale-free networks
  with identical infectivity.
\newblock {\em Physical Review E}, 74:056109.

\end{thebibliography}
\end{document}